\PassOptionsToPackage{hyphens}{url}

\documentclass[sigconf,10pt,authorversion]{acmart}

\usepackage[utf8]{inputenc}
\usepackage[english]{babel}
\usepackage{footnote}
\usepackage{paralist}
\usepackage[capitalise]{cleveref}
\usepackage[acronym]{glossaries}
\usepackage{listings}
\newcommand{\code}[1]{\lstinline{#1}}

\copyrightyear{2018}
\acmYear{2018}
\setcopyright{rightsretained} 
\acmConference[MobiCom '18]{The 24th Annual International Conference on Mobile Computing and Networking}{October 29--November 2, 2018}{New Delhi, India}
\acmBooktitle{The 24th Annual International Conference on Mobile Computing and Networking (MobiCom '18), October 29--November 2, 2018, New Delhi, India}
\acmPrice{15.00}
\acmDOI{10.1145/3241539.3267716}
\acmISBN{978-1-4503-5903-0/18/10}

\newacronym{AWDL}{AWDL}{Apple Wireless Direct Link}
\newacronym{AW}{AW}{Availability Window}
\newacronym{TLV}{TLV}{Type-Length-Value}
\newacronym{AF}{AF}{Action Frame}

\begin{document}

\title[Demo: Cross-Platform Ad hoc Communication with AWDL]{Demo: Linux Goes Apple Picking: Cross-Platform Ad hoc Communication with Apple Wireless Direct Link}

\author{Milan Stute}
\orcid{0000-0003-4921-8476}
\affiliation[obeypunctuation=true]{
	\department{Secure Mobile Networking Lab}\\
	\institution{TU Darmstadt},
	\country{Germany}
}
\email{mstute@seemoo.de}

\author{David Kreitschmann}
\affiliation[obeypunctuation=true]{
	\department{Secure Mobile Networking Lab}\\
	\institution{TU Darmstadt},
	\country{Germany}
}
\email{dkreitschmann@seemoo.de}

\author{Matthias Hollick}
\affiliation[obeypunctuation=true]{
	\department{Secure Mobile Networking Lab}\\
	\institution{TU Darmstadt},
	\country{Germany}
}
\email{mhollick@seemoo.de}

\begin{abstract}

\gls{AWDL} is a proprietary and undocumented wireless ad hoc protocol that Apple introduced around 2014 and which is the base for applications such as AirDrop and AirPlay.
We have reverse engineered the protocol and explain its frame format and operation in our MobiCom~'18 paper ``One Billion Apples' Secret Sauce: Recipe of the Apple Wireless Direct Link Ad hoc Protocol.''
\gls{AWDL} builds on the IEEE\,802.11 standard and implements election, synchronization, and channel hopping mechanisms on top of it. Furthermore, \gls{AWDL} features an IPv6-based data path which enables direct communication.

To validate our own work, we implement a working prototype of AWDL on Linux-based systems. Our implementation is written in C, runs in userspace, and makes use of Linux's \emph{Netlink} API for interactions with the system's networking stack and the \emph{pcap} library for frame injection and reception.
In our demonstrator, we show how our Linux system synchronizes to an existing \gls{AWDL} cluster or takes over the master role itself. Furthermore, it can receive data frames from and send them to a MacBook or iPhone via \gls{AWDL}. We demonstrate the data exchange via ICMPv6 echo request and replies as well as sending and receiving data over a TCP connection.

\end{abstract}



\maketitle


\section{Introduction and Background}

New types of proximity-based services such as contactless payment (e.\,g., NFC), location-aware advertisements (e.\,g., Bluetooth LE), user-aware security measures (e.\,g., Apple Auto Unlock), peer-to-peer file transfers (e.\,g. Apple AirDrop), and media streaming (e.\,g., Apple AirPlay) have re-ignited the interest in wireless ad hoc communications.
One key enabling technologies for such services is \gls{AWDL} which provides Wi-Fi speed data transfers between neighboring devices but is---unfortunately---only available in Apple devices.
In our MobiCom'18 paper~\cite{Stute2018}, we investigated the workings of this protocol, released an open source Wireshark dissector~\cite{awdl-wireshark}, and conducted a performance evaluation with Apple's implementations.
In this paper, we draw on these findings and present a working prototype of \gls{AWDL} as a Linux userspace daemon which can take part in the \gls{AWDL} election and synchronization process, can be discovered by others, and receives and transmits data frames from and to other \gls{AWDL} devices. Our implementation integrates itself in the Linux networking stack by providing a virtual network interface such that existing IPv6-capable programs can use \gls{AWDL} without modification.
Our setup consists of a Linux-based machine and macOS/iOS devices, and we can show that arbitrary programs (e.\,g., \code{ping} and \code{netcat}) successfully run over our \gls{AWDL} implementation.
Our work proves that cross-platform ad hoc communication is feasible and we provide a base for future cross-platform ad hoc applications.


\section{Implementation}

We implement our prototype in plain C for performance reasons and to facilitate porting the code to other platforms. 

\begin{figure}
	\includegraphics[width=\linewidth]{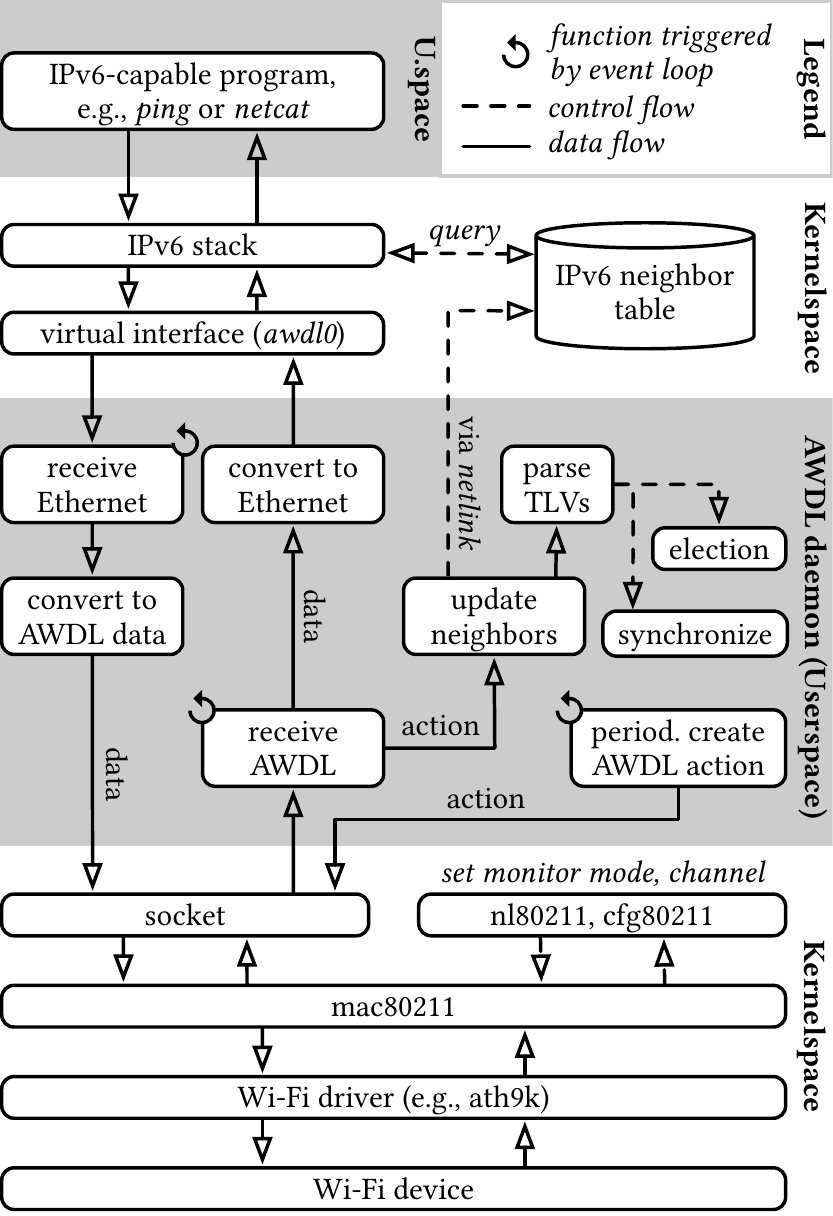}
	\caption{Architecture of our \gls{AWDL} prototype and integration in the Linux networking stack.}
	\label{fig:impl}
\end{figure}

\subsection{Overview}

We depict the architecture and integration of our \gls{AWDL} daemon in \cref{fig:impl}.
At its core, the daemon uses an event loop (\emph{libuv}\footnote{\url{http://libuv.org}}) that
\begin{inparaenum}
	\item listens on the Wi-Fi interface for new IEEE\,802.11 frames using \emph{libpcap},\footnote{\url{https://github.com/the-tcpdump-group/libpcap}}
	\item listens on a virtual Ethernet interface for new traffic from the host system, and
	\item periodically schedules the transmission of \gls{AWDL} action frames that carry information used for peer discovery, synchronization, and election procedures.
\end{inparaenum}

When a new Wi-Fi frame is available on the monitoring interface, we check whether the frame is an \gls{AWDL} action or data frame. Other frames such as regular IEEE\,802.11 frames are dropped by a BPF filter that only forwards frames with the \gls{AWDL}-specific BSSID \lstinline{00:25:00:ff:94:73}.
If we receive an action frame, we derive~\cite{RFC4291} a link-local IPv6 address from the source Ethernet address and add both to the system's neighbor table. Based on the included \gls{TLV} fields, we run the election and synchronization mechanisms as described in~\cite{Stute2018}.
If we receive a data frame, we strip the \gls{AWDL} data header, replace it with a regular Ethernet header, and forward the frame to the virtual \lstinline{awdl0} interface. We do the inverse for Ethernet frames that we receive from \lstinline{awdl0} and add the \gls{AWDL} sequence number from an internal counter.
In addition, the daemon periodically emits \gls{AWDL} action frames that it builds from its internal synchronization and election state. We document the complete frame format in our Wireshark dissector~\cite{awdl-wireshark}.

\subsection{Portability and Future Work}

Since our prototype is written in C, it should be possible to port the code to different operating systems. However, we have the following dependencies that each target platform needs to provide:
\begin{inparaenum}
	\item a Wi-Fi card supporting active monitor mode with frame injection to be able to receive and send IEEE\,802.11 frames,
	\item a means to change the Wi-Fi channel such as \emph{nl80211},
	\item access to the system's IPv6 neighbor table, and
	\item a facility to create virtual network interfaces such as TUN/TAP.
\end{inparaenum}
In principle, this should allow implementations on Android smartphones where monitor mode and frame injection can be enabled using the Nexmon framework~\cite{nexmon:project}.
Our prototype currently lacks a channel switching mechanism that would be required to follow nodes to a different channel. However, since \gls{AWDL} devices usually meet on one social channel (6, 44, or 149), our prototype still works by continually listening on a fixed channel.

\subsection{Enabling \gls{AWDL} in macOS and iOS Third-party Applications}

On Linux, every program using sockets can use our \lstinline{awdl0} interface.
On macOS, programs must set an XNU-specific \lstinline{SO_RECV_ANYIF}\footnote{\url{https://opensource.apple.com/source/xnu/xnu-4570.41.2/bsd/sys/socket.h}} socket option.\footnote{The socket option is the ``default packet filter'' that we discuss in~\cite{Stute2018}.}
Using this option, any software on macOS using sockets can support \gls{AWDL}, thus, enabling cross-platform applications with minor modifications to the code.
As an alternative, programs on macOS and iOS can use the higher-level \lstinline{NetService} API~\cite{Apple:NSNetServiceRef} which activates mDNS/DNS-SD and establishes TCP connections via the \lstinline{awdl0} interface. We provide an example application implementing a TCP--\gls{AWDL} proxy~\cite{proxawdl}. For cross-platform communication, the Linux system must support mDNS/DNS-SD, e.\,g., via avahi.\footnote{\url{https://www.avahi.org}}


\section{Demonstrator}

\begin{figure}
	\includegraphics[width=\linewidth]{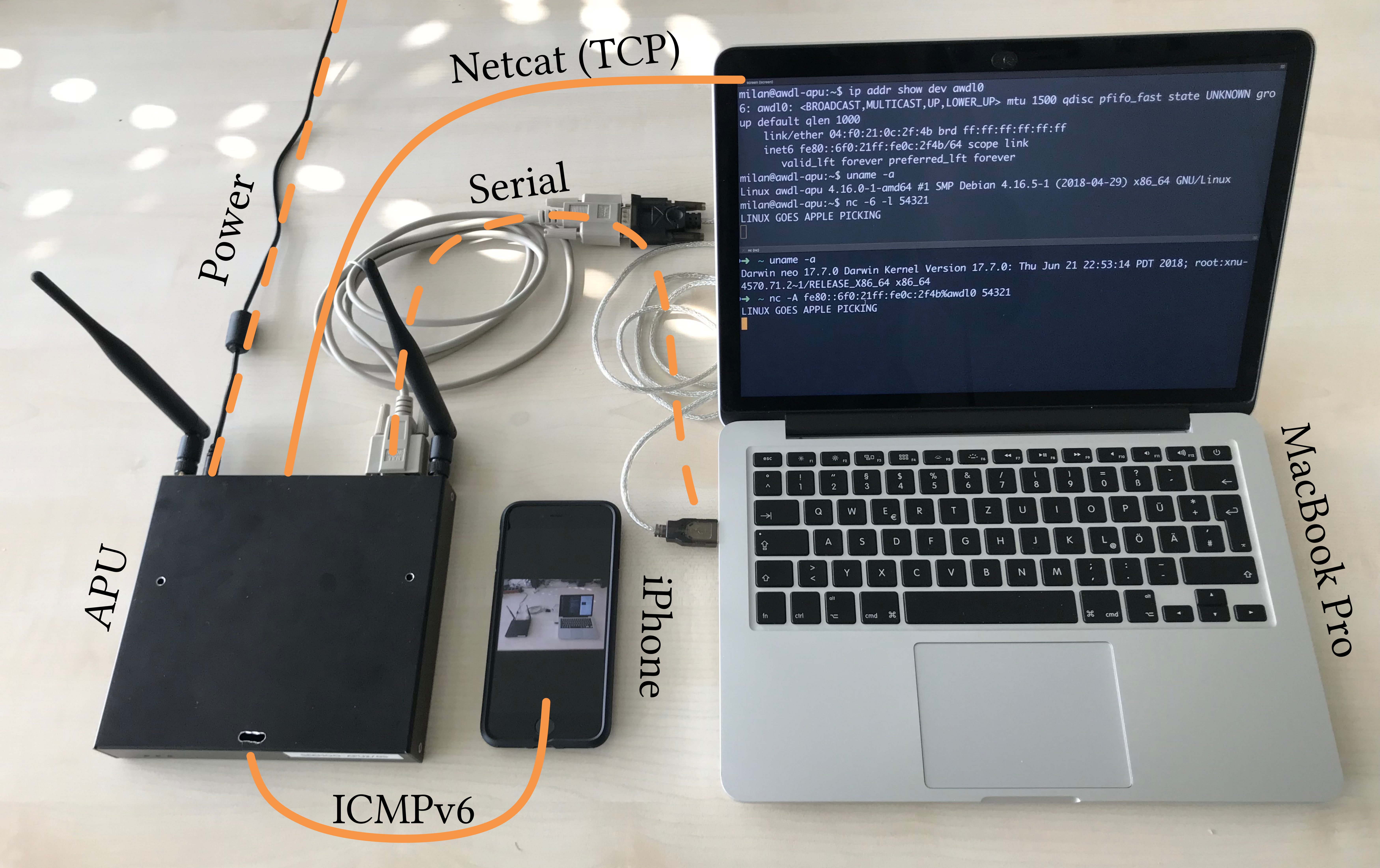}
	\caption{Demonstrator setup consisting of a Linux-based APU board, an iPhone 8, and a MacBook Pro. The terminal on the MacBook's screen shows a working TCP-over-\gls{AWDL} connection between the APU board and the MacBook.}
	\label{fig:setup}
\end{figure}

We briefly summarize our demonstrator devices, the activities that the attendee can see, and list our requirements for our on-location setup.
We show our setup in \cref{fig:setup}.

\subsection{Devices}

Our \gls{AWDL} implementation runs on an APU board~\cite{APU} with a Qualcomm Atheros AR928X Wi-Fi card which implements the IEEE\,802.11n standard and uses the ath9k driver which provides frame injection in monitor mode.
We use various Apple devices such as iPhone and MacBook to demonstrate cross-platform communication.

\subsection{Activities by Attendee}

\begin{figure}
	\includegraphics[width=\linewidth]{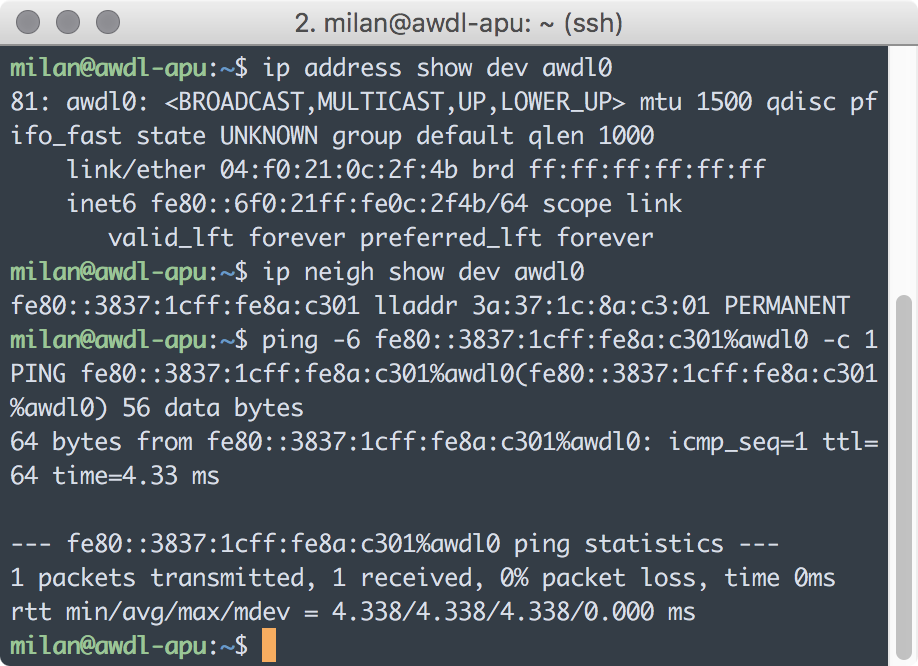}
	\caption{Terminal showing our \gls{AWDL} implementation running on a Linux host.}
	\label{fig:awdl-terminal-linux}
\end{figure}

We can offer to show different aspects of our cross-platform communication depending on the attendees' demands.
\begin{itemize}
	\item We can show the neighbor tables in both Apple and Linux implementations which contains the \gls{AWDL} peers if they emit action frames.
	\item We can conduct a Wireshark live capture of \gls{AWDL} frames and dissect their content. We can then also analyze these capture files w.r.t.\ election behavior and synchronization accuracy similar to~\cite{Stute2018}.
	\item We can send and receive ICMPv6 echo requests and replies by using \code{ping}.
	\item We can establish a TCP connection between two \gls{AWDL} nodes and send messages by using \code{netcat}.
	\item We invite attendees to do all of the above with their own macOS or iOS devices as well.
\end{itemize}
We depict some of the above in \cref{fig:awdl-terminal-linux} and might offer more activities subject to the results of our ongoing research.

\subsection{Requirements at Location}

For our demonstrator, we require
\begin{inparaenum}[(1)]
	\item a table,
	\item three power outlets with Europlugs next to the table, and
	\item about one hour for the setup.
\end{inparaenum}

\begin{acks}
This work is funded by the LOEWE initiative (Hesse, Germany) within the NICER project and by the German Federal Ministry of Education and Research (BMBF) and the State of Hesse within CRISP-DA.
\end{acks}

\bibliographystyle{ACM-Reference-Format}
\bibliography{bibexport}

\end{document}